# Ultrafast Switching in Terahertz Metamaterials using Ion Implanted Silicon on Sapphire

Dibakar Roy Chowdhury[*,1,2], Ranjan Singh[1,3,4], Antoinette J. Taylor[1] and Abul K. Azad[1]

[1]*Center for Integrated Nanotechnologies, Materials Physics and Applications Division, Los Alamos National Laboratory, Los Alamos, New Mexico 87545, USA*

[2]*Center for Sustainable Energy Systems, College of Engineering and Computer Science, Australian National University, Canberra, 0200, Australia*

[3]*Division of Physics and Applied Physics, School of Physical and Mathematical Sciences, Nanyang Technological University, Singapore 637371, Singapore*

[4]*Centre for Disruptive Photonic Technologies, School of Physical and Mathematical Sciences, Nanyang Technological University, Singapore 637371, Singapore*

**Abstract**

We demonstrate ultrafast resonance switching of terahertz metamaterials through optical excitation of radiation damaged silicon placed in the gap of single split gap ring resonator. We observe the dynamic switching OFF of the fundamental resonance mode on a time scale of 4 picoseconds (ps) followed by the switching ON of the same resonance after 20 ps. Electric field distributions in the metamaterials unit cell derived through numerical simulations clearly support our experimental observations, showing that the high electric field at the resonator gaps, responsible for inductive-capacitive resonance (LC), completely disappears and switches OFF the resonance after being optically excited. The ultrafast switching of the metamaterial resonance is attributed to the generation of free carriers in ion-implanted silicon and their recombination at an ultra-short time scale. Such silicon based active control of metamaterials can lead to the ultrafast terahertz metadevices.

[*]Email:dibakarrc@gmail.com

Silicon is always preferable material since it is cost effective and ideal for mass fabrication of electronic and photonic devices [1-3]. The rapid progress in microelectronics industry is made possible mainly due to the abundant availability of silicon. On the other hand, metamaterials are novel artificial materials with unusual electromagnetic properties emerging from their geometrical structure rather than the composition [4-13]. Since metamaterials enable the manipulation of light in exotic ways, they hold the promise for an entire range of novel devices with on-demand functionalities. Additionally, metamaterials are size scalable and much simpler to fabricate at microwave [6, 14], terahertz (THz) [7, 15-19] and optical [8, 20-21] frequencies which makes them even more attractive. In this work, we have integrated ion-implanted silicon in THz metamaterials in order to realize ultrafast switching. The main purpose of this integration is to utilize the benefits of silicon technology and add ultrafast dynamic characteristic to the active metamaterial device in order to exploit the unique electromagnetic properties. Additionally, terahertz regime is an ideal platform to demonstrate these effects since electronics and photonics can merge at this frequency domain. Most metamaterials consist of split ring metallic resonators as the basic building blocks. Operation of metamaterials based devices at user defined frequencies depends on suitable control of fundamental and higher order resonance modes of these split ring resonators (SRR) [9-22]. In this work we demonstrate ultrafast control of intrinsic metamaterial resonances utilizing ion-implanted silicon as the dynamic material. This scheme has potentials to enable ultrafast terahertz devices by combining the emerging phenomena of metamaterials with well-established silicon platforms.

In this study, we demonstrate the ultrafast dynamic modulation of the fundamental inductive capacitive (*LC*) resonance of terahertz split ring resonator (SRR). The metamaterial samples studied in this work consist of single gap SRRs with a radiation damaged silicon layer placed only in the SRR's gap. This scheme allowed us to photo-dope and tunes the silicon conductivity at an ultrafast time scale only in the gap region of the SRR, in contrast to previous schemes where the properties of the entire substrate were modified [23].

The ultrafast metamaterial samples studied in this work were fabricated on ion-implanted silicon-on-sapphire (SOS) wafer. The wafer consists of a 660-nm-thick epilayer of <100> oriented radiation damaged silicon epitaxially grown on a 530-µm-thick sapphire substrate. The samples were fabricated using two steps photolithography. First, photolithography was done to create patterns to form metal resonators. Followed by this step, 10-nm-thick titanium and 200-nm-thick gold layer were deposited by electron beam evaporation, which was then lifted-off to form the single gap SRR array. In the second step, photolithography was performed to cover the split gap of the SRR but to expose the remaining silicon layer across the entire sample area. With plasma etching, silicon was removed from all over the sample except in the gap region of the SRR which was masked by the photo resist. Schematic and images of the fabricated sample are shown in Fig. 1 with the detailed geometric dimensions. The metamaterial samples were characterized using the optical pump terahertz probe (OPTP) measurement technique [24, 25] where the polarization of the incident THz electric field was carefully aligned parallel to the gap bearing SRR arm in order to excite the resonance modes (Fig. 1b). A near-infrared femtosecond (fs) laser beam, with a pulse duration of 30 fs, energy of 3.2 mJ/pulse operating at 800 nm with a 1 kHz repetition rate, was employed for terahertz generation/detection and optically pumping the sample. The pump laser beam had a beam diameter of ~1 cm, much larger than the focused THz spot diameter of ~3 mm at the sample, providing uniform excitation over the SOS film in the split gaps of our metamaterial sample. The time delay between optical-pump and terahertz-probe pulses was precisely controlled using a translation stage to change the path length of the optical pump beam. For several pump-probe time delays, the THz signal was measured in the time domain after transmission through the metamaterial samples. The time domain signals were transformed to the frequency domain and normalized using a bare sapphire substrate as the reference, measured simultaneously. In this experiment all the measurements were performed at room temperature and in a dry atmosphere in order to mitigate water vapour absorption.

The carrier lifetime dynamics of the ion implanted unpatterned SOS film at pump fluence of 1400 mW (1400 µJ/cm$^2$) was measured. Figure 2a depicts change in transmitted THz peak signal as a function of time delay between the optical pump beam and the THz probe pulses. The near-infrared photons excite the photocarriers in the silicon layer, initially causing a decrease in the transmitted THz signal because of the photo-induced changes in the transient conductivity. As the time delay increases, the transmitted THz signal increases due to carrier trapping in the ion implanted silicon [**26**]. Fig. 2a shows that the THz peak is reduced as the optical pulse arrives and then recovers to its full strength within 20 ps (approximately). As revealed by the carrier dynamics of the ion implanted silicon film, the free charge carrier concentration in silicon layer changes as a function of the pump probe time delay (Fig. 2a) due to the characteristic carrier lifetime. This change in charge carrier concentration is directly reflected in change in conductivity as demonstrated in figure 2a. Terahertz transmission measurements through the metasurface sample were carried out at several pump probe delays as indicated in figure 2a.

The amplitude transmission through the metamaterial sample shows strong fundamental resonance at point A when measured before the arrival of the pump pulse (see Fig. 2a). The fundamental resonance is excited at 1.08 THz. At a time delay of ~ 4 ps after the pump pulse (point B in figure 2), we observe the complete disappearance of the fundamental resonance; instead, we see a flat transmission spectra at 1.08 THz where the fundamental resonance previously existed. At a time delay of ~6 ps (point C), the transmission still remains flat without any indication of a resonance. At a longer time delay of ~12 ps (point D in Fig. 2a), there is an evolution of a feeble resonance feature near 1.08 THz. At delay of 24 ps, the strong fundamental resonance reappears again at 1.08 THz (point E in Fig. 2b).

Before the arrival of pump beam (point A), the transmission spectrum of the metamaterial sample shows strong resonance due to the intrinsic dielectric nature of the split gap. Silicon in the split gap behaves like dielectric material and hence can support the fundamental LC resonance mode. This fundamental mode is excited by the incident electric field due to the asymmetry of the SRR [**18, 22, 27**]. As the pump-probe delay increases, the transmission spectra measured at point B, the fundamental resonance disappears completely. This happens because of the conductive nature of the silicon in the split gaps due to photo-excited carrier generation. At point C, the silicon islands inside the split gap still remains conductive hence the resonance remains suppressed. Although at this point, a fraction of the photo-carriers have either recombined or been trapped, the overall conductivity of the silicon still remains high enough to prohibit the electric field build-up required to drive LC oscillations. At still longer delays (point D), a major fraction of photocarriers depart from the conduction band through the carrier trapping and recombination processes. This reduction of the conductivity is confirmed by the change in terahertz transmission measured in Fig. 2a. Due to this reduction in conductivity of the silicon in split gaps we observe the reappearance of the fundamental resonance at 1.08 THz. When the optical pump THz delay is nearly equal to 24 ps (point E), all the excited carriers are completely relaxed restoring the dielectric nature of the silicon region inside the split gap, therefore, the SRR is able to support full-strength LC resonance at 1.08 THz as revealed in Fig. 2b.

In order to understand the intriguing phenomena behind the resonance switching, we performed finite-element simulations using commercially available numerical software, Computer Simulation Technology (CST) [**28**]. In the full wave numerical simulations we have used periodic boundary conditions to simulate a single unit cell, as shown in figure 1b and in the inset of figure 3. The periodic boundary conditions form the SRR array, as seen in figure 1a during the simulations. An adaptive mesh configuration is employed, generating ~30000 tetrahedron mesh cells during the numerical simulations. The transmitted component of the S parameter through the sample is normalized with a blank sapphire substrate. During the simulation the excited E-field was set parallel to the split gap of the resonator as the polarization shown in figure 1b. The metal resonators (gold) were simulated as lossy metal with a conductivity of $6.7 \times 10^7$ S/m, whereas the sapphire substrate was modelled as a lossless dielectric with permittivity equal to 10.5. Silicon in the gap is modelled as lossy silicon with variable conductivity in order to account for the effect of photo-excitation in the silicon region placed inside the split gap at different pump probe delays. The simulation results are shown in figure 3 and excellent agreement is observed both quantitatively and qualitatively when compared to the experimental measurements. Numerical simulation at point A is done by considering intrinsic silicon in the gap region since split gap is not photodoped. In the next simulation, we considered a silicon conductivity value of 10,000 S/m to reproduce the situation at pump probe delay of 4 pico second (point B in figure 2a). Experimental transmission at point C is modelled by assigning a conductivity equal to 4000 S/m. For point D, a relatively small value of silicon conductivity (200 S/m) is assigned in simulations. We have observed a weakened resonance at this point since there is an initial damping of the resonance strength due to Ohmic losses in the silicon in the gap region (Fig 2b and Fig. 3). At point E, the carriers are completely relaxed from the excited state. So the situation is equivalent to the case we observed at point A in the lifetime dynamics plot. In figure 4, we have plotted the induced electric field distribution for several silicon conductivities but at the same frequency (1.08 THz) where fundamental resonance appeared. With silicon conductivity $\sigma = 0$ S/m (point A), a strong electric field enhancement takes place inside the split gap confirming the mode is due to the inductive capacitive (*LC*) resonance (Fig. 4a). When resonances are measured at the peak of the pump beam, as in the situation at point B of pump-probe delay, the high conductivity of silicon in the split gap is unable to support the fundamental resonance mode. Therefore, we do not see any electric field confinement in the split gap (Fig. 4b). This indicates the complete switch OFF of *LC* resonance mode. As the time delay is increased further (situation at point C), the conductivity inside the split gap still remains high hence still unable to support the fundamental resonance mode. Therefore the induced electric field strength in the split gap remains very low (Fig. 4c) confirming that the *LC* mode is still OFF. With further delay at point D, a large fraction of charge carriers have recombined, so the electrical conductivity of the split gap is reduced considerably and we see the evolution of resonance with reduced strength in transmission. The simulated electric field supports this fact with reduced amount of electric field strength induced in the split gap region as seen in Figure 4d.

In summary, we have demonstrated ultrafast resonance switching of single gap SRR based planar THz metamaterial. Resonance switching on an ultrafast time scale is achieved by selective, but transient, excitation of split gap conductivity. The evolution of the fundamental resonance mode is studied for several pump-probe time delays. Numerical simulation of induced electric field distribution inside the split ring resonator confirms our experimental findings. The silicon based device structure adopted in this work can provide a unique platform to integrate current metamaterials with silicon technology. Additionally this scheme can be helpful in realizing ultrafast metamaterials based THz switches.

We acknowledge support from the Los Alamos National Laboratory LDRD Program. This work was performed, in part, at the Center for Integrated Nanotechnologies, a US Department of Energy, Office of Basic Energy Sciences Nanoscale Science Research Centre operated

**Figure Captions**

FIG. 1 (a) Schematic of the SRR array depicting the MM sample studied in this work. In the image, $P_x$ and $P_y$ are periodicity in the x and y directions, respectively. Optical image of the unit cell of the fabricated MM sample is shown in figure b. The geometrical dimensions of the SRR are indicated inside the figure along with the polarization of the terahertz probe beam. Silicon is placed inside the split gap.

FIG. 2 Carrier lifetime dynamics of ultrafast silicon is shown in figure a. In b, measurements of the transmitted E-field amplitude through the metamaterial sample at several pump-probe delays are shown. In the measurements electric field of polarization is parallel to the SRR split gap.

FIG. 3 Numerical simulations of electric field amplitudes transmitted through the MM sample for different silicon conductivities in the split gap.

FIG. 4 Simulation of induced electric field strength at fundamental resonance frequency (1.08 THz) for different conductivity of silicon in the split gap resembling the situations with different pump probe time delay. In all the subfigures same scale is employed to represent the induced electric field strength.

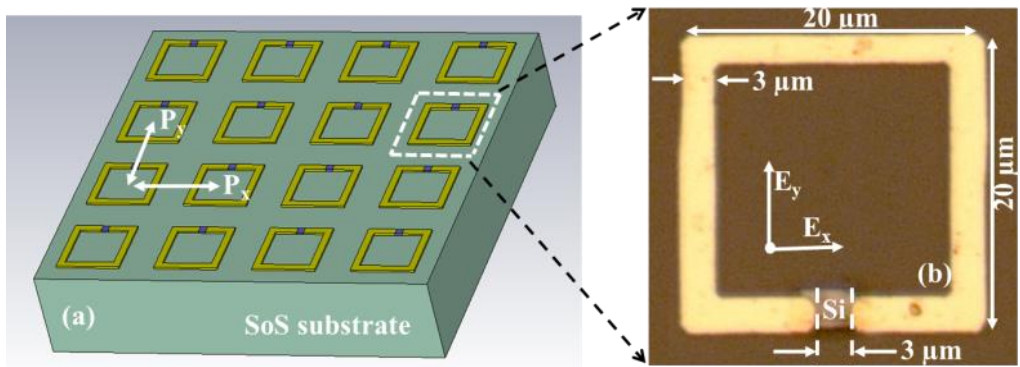

**Figure 1.**
**Roy Chowdhury** *et al.*

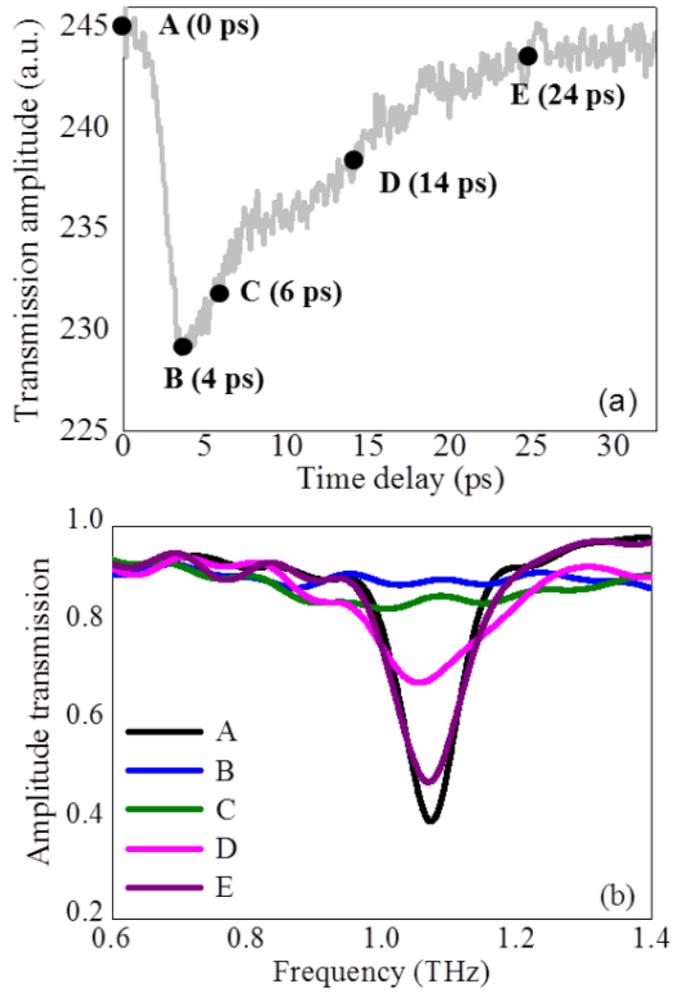

**Figure 2.**
**Roy Chowdhury et al.**

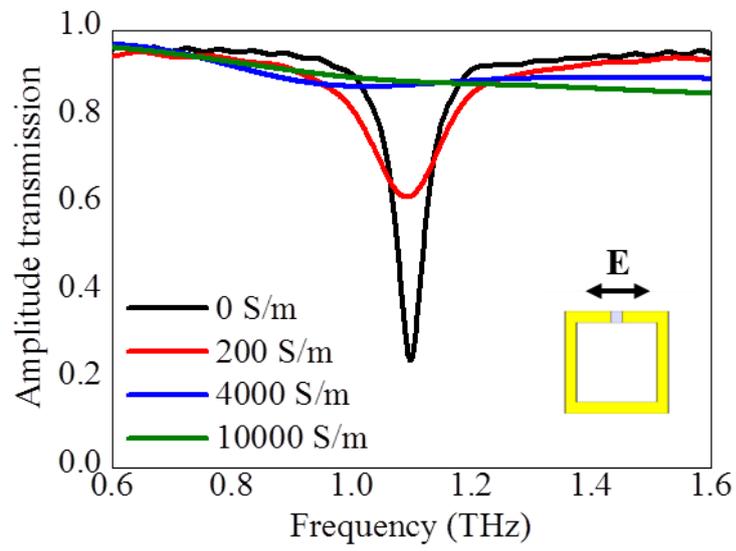

**Figure 3.**
**Roy Chowdhury** *et al.*

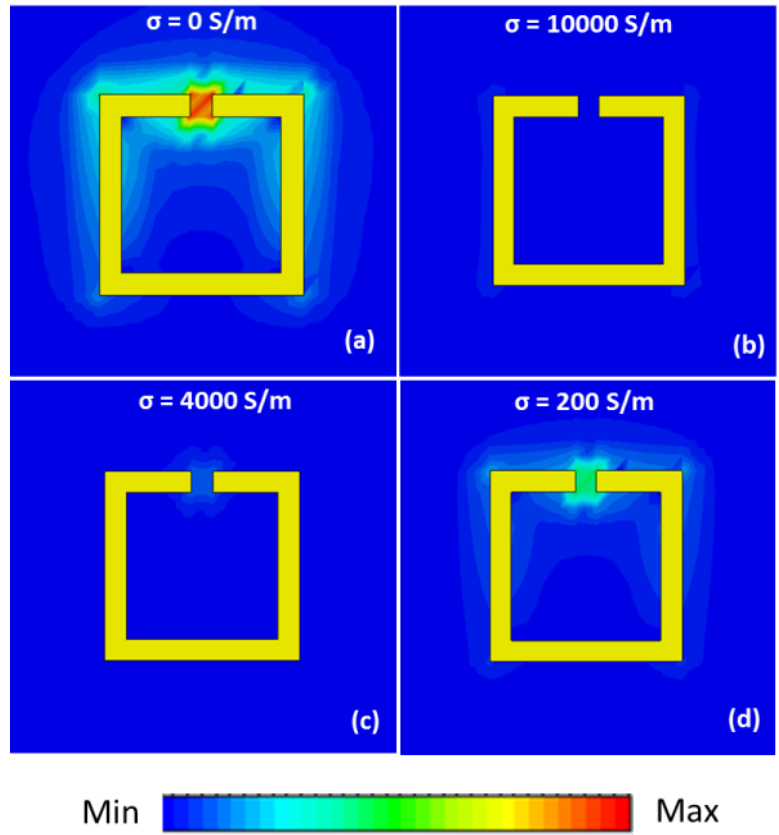

**Figure 4.**
**Roy Chowdhury** *et al.*